# Magnetoresistance and Hall Effect in the Ferromagnetic Semiconductor $Ga_{1-x}Mn_xAs$

**K W Edmonds, R P Campion, K-Y Wang, A C Neumann, B L Gallagher, C T Foxon, P C Main**
School of Physics and Astronomy, University of Nottingham, Nottingham NG7 2RD, UK

*Abstract. The resistivity, temperature, and magnetic field dependence of the anomalous Hall effect in a series of metallic $Ga_{1-x}Mn_xAs$ thin films with $0.015 \leq x \leq 0.08$ is presented. A quadratic dependence of the anomalous Hall resistance on the resistivity is observed, with a magnitude which is in agreement with Berry phase theories of the anomalous Hall effect in dilute magnetic semiconductors.*

In III-V dilute magnetic semiconductors such as $Ga_{1-x}Mn_xAs$, the ferromagnetic interaction between localised spins is mediated by the itinerant valence band electrons [1]. The participation of the free carriers in the magnetism results in a large anomalous Hall effect (AHE). The total Hall resistance is given by

$$\rho_{xy} = \rho^{NH} + \rho^{AH} = B_z/pe + R^A(\rho)M_z. \quad (1)$$

The normal Hall term $\rho^{NH}$ and the anomalous Hall term $\rho^{AH}$ are proportional to the perpendicular components of the external magnetic field, $B_z$, and the magnetisation, $M_z$, respectively. The measured Hall resistance therefore contains information on both the magnetism and the carrier concentration $p$. However, separate extraction of this information is complicated by the field dependence of the resistivity $\rho$ and the anomalous Hall coefficient $R^A(\rho)$, so that obtaining accurate values for the hole density in $Ga_{1-x}Mn_xAs$ requires high magnetic fields and highly metallic samples [1,2]. A better understanding of the factors affecting the AHE is therefore desirable.

The AHE is usually ascribed to a scattering anisotropy induced by the spin-orbit interaction [3]. Skew and side-jump descriptions give $R^A \alpha \rho$ and $R^A \alpha \rho^2$ respectively, where $\rho$ is the longitudinal resistance. The skew scattering picture is usually assumed in $Ga_{1-x}Mn_xAs$ [1], although a quadratic dependence on $\rho$ has been observed for $In_{0.88}Mn_{0.12}As$ [4]. It was recently proposed that the AHE in the III-V ferromagnetic semiconductors is dominated by a disorder independent contribution which is due to a Berry phase acquired by the itinerant electrons, and results in a field-independent anomalous Hall conductivity [5]. In the following, we present measurements of the anomalous Hall effect in a series of metallic $Ga_{1-x}Mn_xAs$ samples as a function of magnetic field and temperature, and compare our results to the predictions of reference [5].

45nm thick $Ga_{1-x}Mn_xAs$ films with $0.015 \leq x \leq 0.08$ were grown on GaAs(001) by low temperature molecular beam epitaxy. The Mn concentration was controlled by varying the calibrated Mn/Ga ratio, and confirmed by x-ray fluorescence and x-ray diffraction measurements. Full details of the growth and structural characterisation are presented elsewhere [6]. In agreement with previous studies, a significant increase in $T_C$ was obtained by annealing at low temperatures (≈180ºC in the present case). This procedure resulted in a $T_C$ of up to 132K for $x$=0.06, as well as a reduction in the resistivity and magnetoresistance. Across the whole range of Mn doping studied, the samples are in the so-called 'metallic' regime, with conductivities in the range $(200-700)\Omega^{-1}cm^{-1}$ at low temperatures [2]. Magnetoresistance and Hall measurements were performed on lithographically fabricated Hall bar samples in magnetic fields $B$ between –16.5T and 16.5T.

Figure 1 shows the magnetoresistance $\rho(B)$ and Hall resistance $R_{xy}(B)=\rho_{xy}/d$, where d is the film thickness, of samples with $x$=0.015, 0.06, 0.08, at a temperature of 0.4K. $\rho(B)$ shows an initial steep rise with increasing $B$ due to a reorientation of the sample magnetisation into the perpendicular direction (anisotropic magnetoresistance), followed by a more gradual decrease beyond $B$≈0.5T. This isotropic negative magnetoresistance has been ascribed to spin-disorder scattering in metallic high conductivity samples, and suppression of localisation in insulating samples [7]. Consistent with the former mechanism, and in agreement with previous studies [1], we find that both the resistivity and the magnetoresistance are largest close to the Curie temperature $T_C$. However, we also observe that the resistivity has a minimum value at around 4K, below which the resistivity and the magnetoresistance begin to increase with decreasing temperature. This is inconsistent with spin-disorder scattering, and suggests that localisation effects also give an important contribution in samples showing metallic behaviour. We also observe a clear correlation between resistivity and magnetoresistance, with the smallest magnetoresistance observed in the highest conductivity samples.

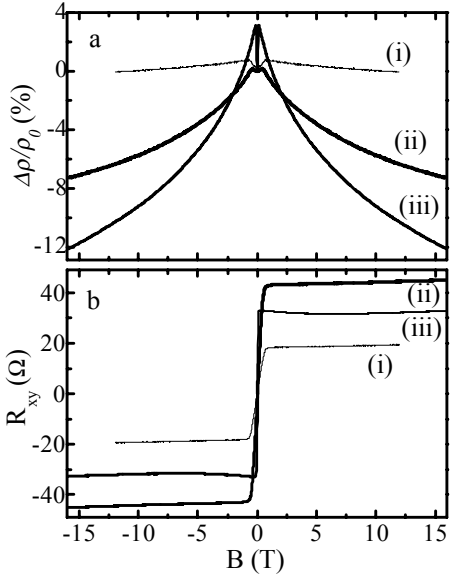

FIGURE 1. (a) magnetoresistance, and (b) Hall resistance, measured at 0.4K for $Ga_{1-x}Mn_xAs$ thin films with (i) $x$=0.06, (ii) $x$=0.08, (iii) $x$=0.015.

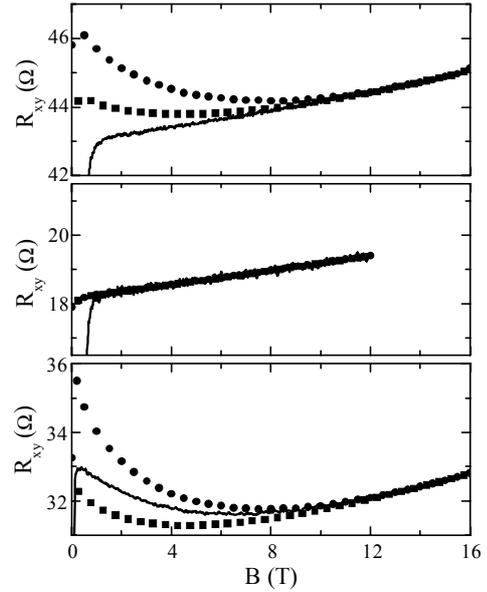

FIGURE 2. Hall resistance for (a) $x$=0.08, (b) $x$=0.06, (c) $x$=0.015 (Full lines); fitted lines using $R^{AH} \alpha \rho$ (squares) and $R^{AH} \alpha \rho^2$ (circles).

The Hall resistance changes rapidly between −0.5T and +0.5T due to the dominant anomalous Hall effect which is proportional to $M_z$. At higher fields, a much more gradual slope is observed. The behaviour at high fields can be seen more clearly in figure 2, where the data is shown on an enlarged scale. For $x$=0.015, the magnitude of $R_{xy}$ falls with increasing $B$ between 0.5T and 6.5T, before monotonically increasing above 6.5T. This is due to competition between the anomalous Hall effect, which decreases with increasing $B$ due to its dependence on $\rho$, and the normal Hall resistance which increases linearly with $B$. Similar behaviour has been reported previously for metallic $Ga_{1-x}Mn_xAs$ samples [1]. For higher values of $x$, the negative gradient of $R_{xy}$ for mid-range values of $B$ is not observed due to the small size of the negative magnetoresistance in these samples. Therefore, the anomalous Hall effect makes only a small contribution to the slope at high magnetic fields, so that hole concentrations can be determined from this with a relatively high degree of accuracy.

The measured Hall resistance was fitted between 14T and 16.5T using equation (1), assuming that the magnetisation is saturated in the perpendicular direction in this field range, and either a linear or quadratic dependence of $R^A$ on $\rho$. The fit therefore contains two free parameters, which give the normal and anomalous contributions to the Hall resistance. The obtained fit is relatively insensitive to the precise field range used between 12T and 16.5T. The fits were then extrapolated down to zero field using the measured field dependence of $\rho$, and the results are compared to the experimental data in figure 2. The extrapolated curves are expected to overestimate the Hall resistance at low magnetic fields due to the incomplete saturation of the magnetisation (the perpendicular direction is the hard magnetic axis). For $x$=0.08, better agreement is obtained with $R^A \alpha \rho$ than for $R^A \alpha \rho^2$ at low fields, while both fits give good agreement at high fields. However, a quadratic dependence on $\rho$ and a more gradual approach to magnetic saturation would be consistent with the measured Hall resistance. For $x$=0.06, the two fits give almost identical results due to the very small magnetoresistance for this sample (<1% at 10T). For $x$=0.015, the measured Hall resistance cannot be reproduced if $R^A$ is purely linear with $\rho$; the extrapolated curve underestimates the measurement over the range 0.2T to 10T, implying that the magnetisation decreases with increasing field over this range, which is an unphysical result. We have been unable to produce a realistic fit to the experiment data for this sample if $R^A$ is purely proportional to $\rho$, indicating that there is an important $\rho^2$ contribution to the anomalous Hall effect, at least for $x$=0.015.

Varying the temperature gives another means of studying the resistivity dependence of the anomalous Hall effect, as even the highest conductivity samples show an increase in resistivity with decreasing temperature below around 4K. For the sample with $x$=0.06, the very small magnetoresistance allows normal and anomalous components to be separated with a high degree of accuracy, so that the temperature dependence of the two components can be studied. Furthermore, because of the high $T_C$ of this sample (125K), we can neglect variations of the saturation magnetisation by restricting the study to the temperature range 0.4K to 5K. The normal Hall effect (and thus the hole density) showed no temperature dependence within the experimental error, while the anomalous Hall effect showed a clear dependence on the resistivity. The anomalous Hall resistance is plotted in figure 3 versus the resistivity for fixed $B$=10T and varying temperature. Best fit lines for $R_{xy} \alpha \rho$ and $R_{xy} \alpha \rho^2$ are shown. The data is in good agreement with the quadratic fit. Behaviour consistent with this is observed in other samples studied. The exception is the sample with $x$=0.08, for which the Hall resistance is observed to vary approximately as $R_{xy} \alpha \rho^{1.4}$, indicating that linear and quadratic contributions are both important for this sample.

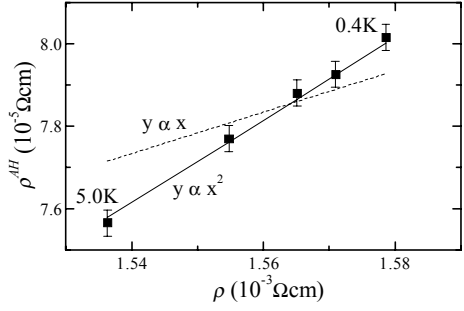

FIGURE 3. anomalous Hall resistivity $\rho^{AH}$ versus longitudinal resistivity $\rho$ at B=10T and varying T between 0.4K and 5.0K for an annealed $x$=0.06 sample (squares); best fit lines assuming $R^{AH} \alpha \rho$ (broken line) and $R^{AH} \alpha \rho^2$ (full line).

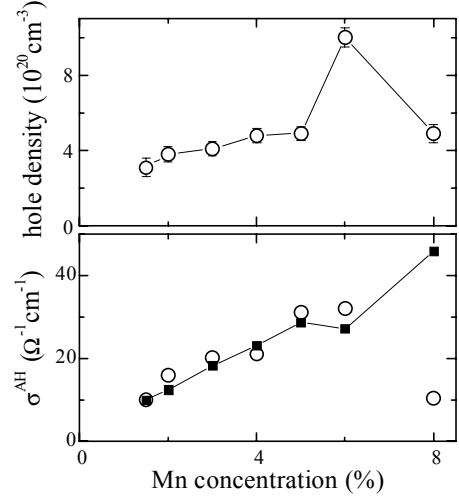

FIGURE 4. (a) hole density and (b) anomalous Hall conductivity versus Mn concentration $x$ for a series of $Ga_{1-x}Mn_xAs$ samples. Open circles in (b) are the experimental result, closed squares are the predicted values using equation (6) of ref. [4].

The observed quadratic dependence on $\rho$ is equivalent to a constant anomalous Hall conductivity, $\sigma^{AH} \approx \rho^{AH}/\rho^2$, as predicted in ref. [5]. Equation (6) in ref. [5] gives a estimate for the upper bound of the Berry phase related AHE, in the limit of large spin-orbit coupling and $m_{hh}>>m_{lh}$, as a function of the hole and Mn concentrations and the p-d exchange coupling strength $J_{pd}$. The hole densities for the present samples are obtained from the fits to the Hall effect data, and are shown in figure 4a. The variation of hole density with $x$ is discussed in more detail elsewhere [2]. For $J_{pd}$ we use a value of 50meV nm$^{-3}$, consistent with ref. [5] and experimental observations [1]. The experimental values of $\sigma^{AH}$ are plotted together with the theoretical upper bound in figure 4b. With the exception of the $x$=0.08 sample, all of the experimental results are comparable with theory. This good agreement should be should be taken with caution, since the calculation assumes that the spin-orbit coupling strength is infinite, and neglects band warping; a more realistic treatment tends to predict smaller values of $\sigma^{AH}$ [5]. However, the magnitude of the measured anomalous Hall effect is consistent with the prediction that there is an important contribution to the measured AHE due to a disorder-independent Berry phase term, while not ruling out additional scattering dependent contributions.

In conclusion, by performing a fit to measured Hall resistance data, normal and anomalous contributions to the Hall effect in $Ga_{1-x}Mn_xAs$ are extracted. This allows the determination of hole densities with a relatively small experimental error. From the temperature dependence of the anomalous Hall resistance, we observe a quadratic dependence on the resistivity which is consistent with the prediction of a scattering independent anomalous Hall conductivity in dilute magnetic semiconductors.